# Monte Carlo modeling of the effect of *extreme events* on the extinction dynamics of animal species with 2-year life cycles

**Sukanto Bhattacharya**
Alaska Pacific University

**Sreepurna Malakar**
University of Alaska

## Abstract

Our paper computationally explores the *extinction dynamics* of an animal species effected by a sudden spike in mortality due to an *extreme event*. In our study, the animal species has a 2-year life cycle and is endowed with a high survival probability under *normal* circumstances. Our proposed approach does not involve any restraining assumptions concerning environmental variables or predator-prey relationships. Rather it is based on the simple premise that if observed on an year-to-year basis, the population size will be noted to either have gone up or come down as compared to last year. The conceptualization is borrowed from the theory of asset pricing in stochastic finance. The survival probability ($\lambda$) is set at unity i.e. the model assumes that all young members of the population mature into adults capable of reproduction. As we bias our model heavily in favor of survival, the chance of the population size increasing over time is much higher than it suffering a decline, if no extreme events occur. One of the critical parameters in our simulation model is the *shock size* i.e. the maximum number of immediate mortalities that may be caused by an extreme event. We run our model for two pre-selected fecundity levels denoted as "high" and "low". Under each of the two fecundity levels one hundred independent simulation runs are conducted over a time period of ten years (i.e. five generations) and the relevant descriptive statistics are reported for the terminal (i.e. the fifth) generation. Shock sizes are varied until at least one scenario of total extinction is observed in the simulation output. Any extinctions occurring in $t_0$ are treated as "trivial cases" and not counted. Our results indicate that an extreme event with a maximum shock size exceeding $2/3$ the size of the pristine population can potentially drive any animal species with a 2-year life cycle to extinction for both "low" and "high" fecundity levels.

---

## Background and research objective

The primal problem which proponents of *population viability analysis* (PVA) initially set out to solve concerned the minimum size of population of a species that is required for it to have a *reasonable probability of survival* over a *reasonable length of time*. [1]

However, animal species with no clearly observed threat to either their population size or fecundity may nevertheless become threatened with extinction within a relatively short period of time as result of a sudden, massive spurt in mortality due to man-caused disasters like oil spills, radiation leakage from nuclear power plants etc. Given the complexity of the problem itself, leave alone its interaction with other complex problems, it is well nigh impossible to come up with an "all-encompassing" equation that could predict the lifespan of any species incorporating the likelihood of surviving even a moderate-scale disaster with an acceptably narrow confidence interval of prediction.

PVA approaches often start off with either *age-specific* or *stage-specific* deterministic survival models and then add on stochastic components along the way as necessitated by the particular population being studied. Although the age-specific models have a longer history in terms of actual applications, the stage-specific models are technically superior mainly because by focusing on life-cycle stages; these models help to focus attention on identifying the critical transitions that may offer appropriate intervention opportunities. However PVA approaches do not normally consider the risk of catastrophic events under the pretext that no population size can be large enough to guarantee survival of a species in the event of a large-scale natural catastrophe. [2] Nevertheless, it is only very intuitive that some species are more "delicate" than others; and although not presently under any clearly observed threat, could become threatened with extinction very quickly if an *extreme event* was to occur even on a low-to-moderate scale. The term "extreme event" is preferred to "catastrophe" because catastrophe usually implies a natural event whereas; quite clearly; the chance of man-caused extreme events poses a much greater threat at present to a number of animal species as compared to any large-scale natural catastrophe.

This paper specifically deals with the population dynamics of animal species that have a two-stage life cycle. However, we believe that the conceptualization can be easily extended to species with n-stage life cycles; where n > 2. An animal has a two-stage life cycle when; in the first stage, newborns become immature youths and in the second stage; the immature youths become mature adults. A number of insect and fish species, the pacific pink salmon (*Oncorhynchus gorbuscha*) being perhaps one of the most widely studied among them, have two-stage life cycles; [3] each stage corresponding to a year. Therefore, in terms of the stage-specific approach, if $Y_t$ denotes the number of immature young in year t and $A_t$ denotes the number of mature adults, then the number of adults in year t + 1 will be some proportion of the young, specifically those that survive to the next (reproductive) stage. Then the formal relationship between the number of mature adults in the next year and the number of immature youths this year may be written as follows:

$$A_{t+1} = \lambda Y_t$$

Here $\lambda$ is the survival probability, i.e. it is the probability of survival of a youth to maturity. The number of young next year will depend on the number of adults this year:

$$Y_{t+1} = f(A_t)$$

Here f describes the reproduction relation between mature adults and next year's young.

This is a straightforward system of simultaneous difference equations which may be analytically solved using a variation of the *cobwebbing approach*. [4] The solution process begins with an initial point ($Y_1$, $A_1$) and iteratively determines the next point ($Y_2$, $A_2$). If *predator satiation* is built into the process, then we simply end up with Ricker's model:

$$Y_{t+1} = rA_t e^{-A_t/K}$$

Here *r* is the maximum reproduction rate (for an initial small population) and K is the population size at which the reproduction rate is approximately half its maximum [5].

It has been shown that if ($Y_0$, $A_0$) lies within the first of three possible ranges, ($Y_n$, $A_n$) approaches (0, 0) in successive years and the population becomes extinct. If ($Y_0$, $A_0$) lies within the third range then ($Y_n$, $A_n$) equilibrate to a steady-state value of ($Y^*$, $A^*$). Populations that begin with ($Y_0$, $A_0$) within the second range oscillate between ($Y^*$, 0) and (0, $A^*$). Such alternating behavior indicates one of the year classes, or cohorts, become extinct while the other persists i.e. adult breeding stock appear only every other year. Thus the model reveals that three quite different results occur depending initially only on the starting sizes of the population and its distribution among the two stages. [6]

We use the same basic model in our research but instead of analytically solving the system of difference equations, we use the same to simulate the population dynamics as a stochastic process implemented on an MS-Excel spreadsheet. Rather than using a closed-form equation like Ricker's model to represent the functional relationship between $Y_{t+1}$ and $A_t$, we use a Monte Carlo method to simulate the stage-transition process with a view to computationally explore the extinction dynamics of an animal species with a 2-stage life cycle; effected by a sudden spike in mortality due to an extreme event. To isolate the effect of extreme event on the extinction dynamics, we heavily bias our model in favor of survival under *normal* circumstances, i.e. the chance of the population size going up is purposely kept much higher than it suffering a decline when *no* extreme events occur.

**Conceptual framework**

In our present model the survival probability is purposely assigned a value of unity. This loads our model heavily in favour of survival i.e. we endow the species with a property that all the young can mature into reproductive adults with a hundred percent certainty. We do this to isolate the effect of the *shock size* (i.e. the maximum number of immediate mortalities) of an extreme event on the extinction dynamics of the species. Setting λ = 1 virtually rules out natural extinction of the species.

We have a chosen a stochastic functional relationship between $Y_{t+1}$ and $A_t$. Moreover the gamma distribution has been chosen to make this stochastic relationship a skewed one. The purpose of imparting the skew is to further weigh the model in favor of survival. Instead of analytically solving the system of simultaneous difference equations iteratively in some variation of the cobwebbing method, we have used them in a spreadsheet model to simulate the population over a time period of ten years (i.e. over five generations).

Our simulation methodology is conceptually borrowed from similar approaches applied in modeling financial systems; mainly in derivative asset pricing models. [7] A financial asset with a starting price of $P_0$ in period $t_0$ is assumed, under a stochastic price evolution process, to either go up to $P_1^+ = P_0 + \Delta P$ or come down to $P_1^- = P_0 - \Delta P$ in $t_1$. N number of simulation runs yield N different possible price paths of the asset and the mean price and variance may be then estimated from the simulation results using the usual statistical methods. A number of *variance reduction techniques* are often applied to tighten the confidence interval estimate of the mean asset price. [8] We apply a similar computational methodology in constructing our model where the initial number of immature young is conceptualized as the starting "asset value". This initial "asset value" goes up or down at the end of every year. If it goes to zero in any year then of course; total extinction occurs.

**Model building**

It is assumed that the initial population consists of one hundred immature young. That is, in $t_0$, there are $Y_0 = 100$ immature young and $A_0 = 0$ mature adults. In the next year i.e. $t_1$, there will be $A_1 = 100$ mature adults (assuming $\lambda = 1$) and these will in turn a new breed of immature young produce $f(A_t)$ before themselves dying out by end of the year.

On lines of a derivative asset pricing problem we perceive that $Y_{t+1} = f(A_t)$ can either *go up* or *come down* at each sample point (year); similar to the *two-state option pricing model* based on a discretized *geometric Brownian motion*. To simulate a standard Brownian motion, one needs to repeatedly generate independent Gaussian random variables with mean 0 and standard deviation $\sqrt{(1/P)}$ where P is the total number of sample points. [9] However, as we want to bias our model in favor of survival as far as possible, we have generated our random variables from the cumulative distribution function (*cdf*) of the gamma distribution rather than the Gaussian (normal) distribution. Moreover unlike the conventional two-state option pricing model, magnitude of the change $|\Delta Y|$ is not fixed in our model; [10] but is rather obtained as $|\Delta Y| = C_2 g$ if $\Delta Y$ has a positive sign and $|\Delta Y| = C_3 g$ if $\Delta Y$ has a negative sign. $C_2$ and $C_3$ are two critical parameters and *g* is a probability value randomly drawn from the *cdf* of the gamma distribution. In our spreadsheet model, this probability value varies in the range 0.51881 ~ 0.68269. These boundaries are fixed by generating random integers in the range 1 to 100 and using the generated random integer to define the shape and scale parameters of the gamma distribution. The gamma distribution performs better than the normal distribution when the distribution to be matched is highly right-skewed; as is desired in our model. The combination of a large variance and a lower limit at zero makes it unsuitable to fit a normal distribution in such cases. [11] The probability density function of the gamma distribution is as given follows:

$$f(x, \alpha, \beta) = \{\beta^\alpha \Gamma(\alpha)\}^{-1} x^{\alpha-1} e^{-x/\beta} \text{ for } x > 0$$

Here $\alpha > 0$ is the shape parameter and $\beta > 0$ is the scale parameter of the gamma distribution. The cumulative distribution function may be expressed in terms of the *incomplete gamma function* as follows:

$$F(x, \alpha, \beta) = \int_0^x f(u)du = \gamma(\alpha, x/\beta)/\Gamma(\alpha)$$

In our spreadsheet model, we have F ($R$, $R/2$, 2) as our *cdf* of the gamma distribution. Here $R$ is an integer randomly sampled from the range 1 to 100. An interesting statistical result of having these values for x, α and β is that the cumulative gamma distribution value becomes equalized with the value of [1 - $\chi^2$ ($R$)] having $R$ degrees of freedom [12].

There are three critical parameters in our simulation model. The first parameter $C_1$ is basically a "switch" that controls the sign of ΔY. If this switch is in the "on" state, the change will be positive; if the switch is in the "off" state the change will be negative. The value of the first parameter is fixed at 0.5189 (i.e. a value which is very close to the lower boundary value of our gamma probability range 0.51881 ~ 0.68269). This parameter value; in conjunction with the skewness of the gamma distribution; ensures that the switch stays in the "on" state with a probability between 0.95 and 1. The second and third parameters ($C_2$ and $C_3$ respectively) determine the size of the change. The second parameter controls the size of a positive change (corresponding to the "on" state). For a "high" fecundity level this parameter value is set at 100 while for a "low" fecundity level, this parameter value is set at 50. The third parameter controls the size of a negative change i.e. the shock size (corresponding to the "off" state).

### Analysis of simulation results and future research directions

One hundred independent simulation runs are conducted under each of the two fecundity levels and sample means and standard deviations are reported for the terminal Y values. The starting shock size is set at 50% of $Y_0$ and is increased in equal sized steps of 25% until at least one case of total extinction is observed in the simulation output. Summarized numerical and graphical results of the simulation are given in the Appendix.

Under both levels of the fecundity parameter, the first extinction is recorded in the simulation output when the shock size parameter is set equal to the initial population size. However, as already stated, in our model |ΔY| = $C_j$g, for j = 2, 3. Therefore, actually |ΔY| is quite a bit lower than $C_j$ because g varies in the range 0.51881 ~ 0.68269. This downside perturbation of the changes in population size means that the extinction-causing shock size of the extreme event is actually smaller than the initial population size of 100. So our model indicates that an extreme event with a maximum shock size roughly equal to 69% of the initial population size may be good enough to wipe out an animal species with a 2-year life cycle like the pink salmon within its first five generations.

This has obvious implications in planning conservation efforts of species with 2-year life cycles, especially in habitats which have a non-negligible probability of occurrence of a man-caused extreme event. Attempts to maintain balance in predator-prey relations and minimize juvenile mortality may not be adequate measures in themselves to guarantee survival of a species *even when the species is <u>not</u> threatened with a clear and present*

*risk*. There should be an objective, scientific appraisal of the *maximum shock size* to a species that thrives in a habitat that has a non-negligible chance of occurrence of an extreme event. If this maximum shock size is more than $^2/_3$ the estimated size of the pristine population, then the species could go extinct if an extreme event was to occur.

Conceptually speaking, our proposed modeling methodology can be easily extended to species with n-year life cycles where n > 2. In terms of a spreadsheet implementation this would simply involve putting in an additional column (to hold the stage-transition formula) for each additional stage. In our model $(Y_j, A_j)$ columns are contiguously placed in sequence for each setting (j = 0, 1 … 9), as these are the only two stages in the life cycle. If, say, there were five stages in the life cycle of a certain species, one would need to place five columns contiguously in sequence for each setting of the simulation model.

## *References*:

[1] W. F. Morris and D. F. Doak, *Quantitative Conservation Biology: Theory and Practice of Population Viability Analysis*. Sinauer Associates, Sunderland, MA, 2002.

[2] H. Caswell, *Matrix Population Models: Construction, Analysis and Interpretation*. Sinauer Associates, Sunderland, MA, 2001.

[3] S. Watkinson, *Life After Death: the Importance of Salmon Carcasses to British Columbia's Watersheds*, Arctic 53(1), 92-99, 2000.

[4] F. C. Hoppensteadt, *Mathematical Methods of Population Biology*. Cambridge Univ. Press, NY, 1982.

[5] W. E. Ricker, *Stock and recruitment*, J. Fish. Res. Bd. Canada 11, 559-623, 1954.

[6] F. C. Hoppensteadt and C. S. Peskin, *Mathematics in Medicine and the Life Sciences*. Springer-Verlag New York Inc., NY, 1992.

[7] J. C. Hull, *Options, Futures and Other Derivative Securities*. Prentice-Hall, Englewood Cliffs, NJ, 1993.

[8] N. Madras, *Lectures on Monte Carlo Methods*. Fields Institute Monographs, Amer. Math. Soc., Rhode Island, 2002.

[9] R. Rendleman and B. Bartter, *Two State Option Pricing*, J. Fin. 34, 1092-1110, 1979.

[10] S. Benninga, *Financial Modeling*. The MIT Press, MA, 2000.

[11] N. L. Johnson, S. Kotz and N. Balakrishnan, *Continuous Univariate Probability Distributions, (Vol. 1)*. John Wiley & Sons Inc., NY, 1994.

[12] N. D. Wallace, *Computer Generation of Gamma Variates with Non-integral Shape Parameters*, Comm. ACM 17(12), 691-695, 1974.

**Appendix**

$C_2 = 50$ ("Low" fecundity level)

| Table I : Descriptive statistics of fifth generation population sizes | |
|---|---:|
| Mean | 235.09 |
| Standard Error | 1.09 |
| Median | 237.00 |
| Mode | 235.00 |
| Standard Deviation | 10.86 |
| Sample Variance | 117.98 |
| Kurtosis | 18.14 |
| Skewness | -4.21 |
| Range | 66 |
| Minimum | 181 |
| Maximum | 247 |
| Count | 100 |

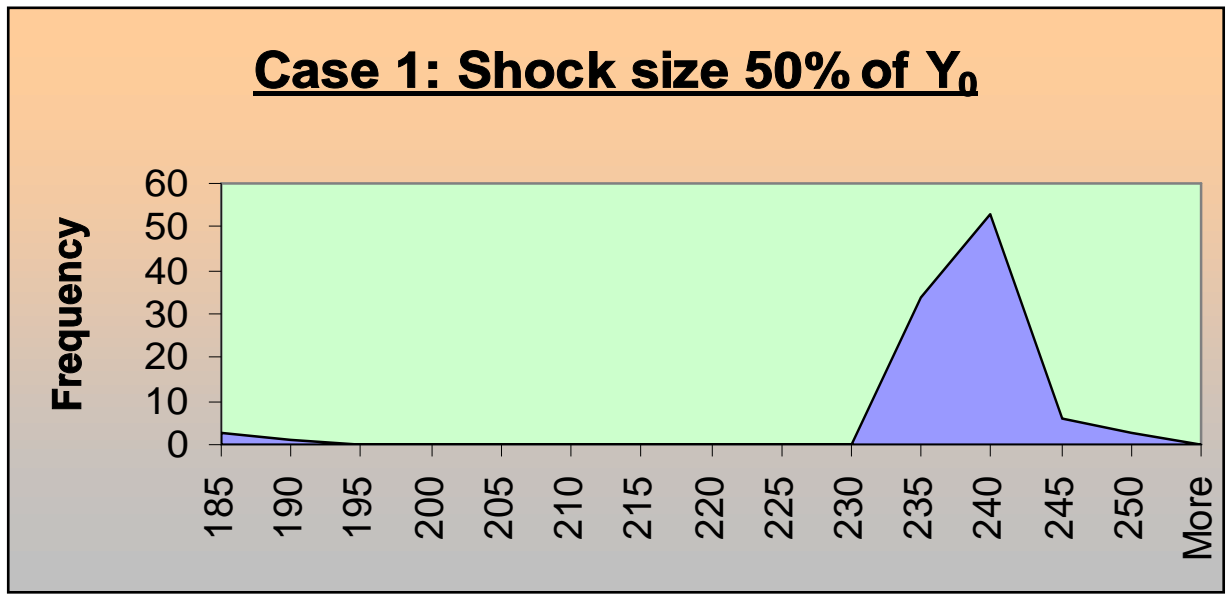

Figure I: Histogram of simulated fifth generation population sizes

| Table II : Descriptive statistics of fifth generation population sizes | |
|---|---:|
| Mean | 234.08 |
| Standard Error | 1.30 |
| Median | 236.00 |
| Mode | 236.00 |
| Standard Deviation | 13.03 |
| Sample Variance | 169.85 |
| Kurtosis | 19.90 |
| Skewness | -4.53 |
| Range | 79 |
| Minimum | 168 |
| Maximum | 247 |
| Count | 100 |

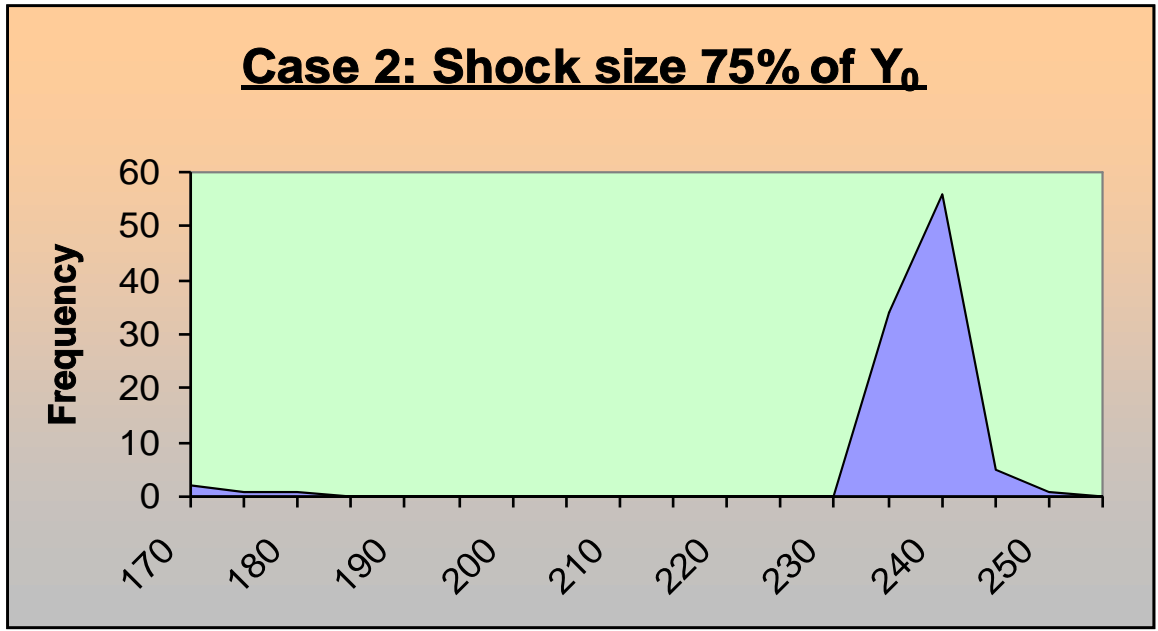

Figure II: Histogram of simulated fifth generation population sizes

| Table III : Descriptive statistics of fifth generation population sizes | |
|---|---:|
| Mean | 226.88 |
| Standard Error | 3.30 |
| Median | 236.00 |
| Mode | 235.00 |
| Standard Deviation | 32.99 |
| Sample Variance | 1088.51 |
| Kurtosis | 22.87 |
| Skewness | -4.22 |
| Range | 250 |
| Minimum | 0 |
| Maximum | 250 |
| Count | 100 |

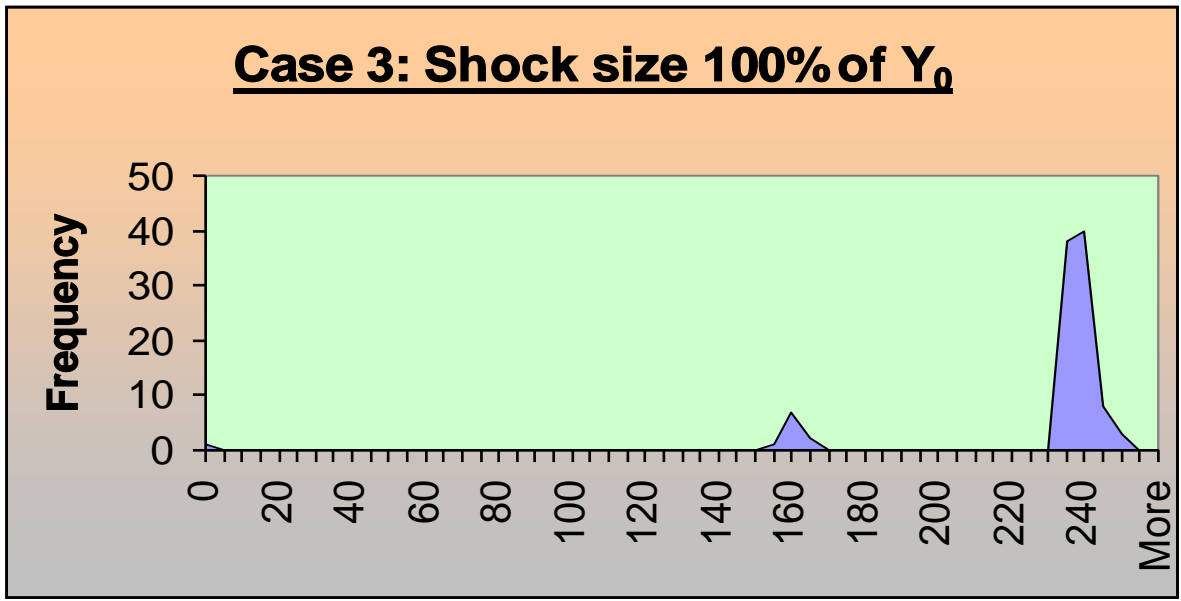

Figure III: Histogram of simulated fifth generation population sizes

$C_2 = 100$ ("High" fecundity level)

| Table IV : Descriptive statistics of fifth generation population sizes | |
|---|---:|
| Mean | 364.87 |
| Standard Error | 2.37 |
| Median | 369.50 |
| Mode | 370.00 |
| Standard Deviation | 23.74 |
| Sample Variance | 563.77 |
| Kurtosis | 6.77 |
| Skewness | -2.74 |
| Range | 111 |
| Minimum | 283 |
| Maximum | 394 |
| Count | 100 |

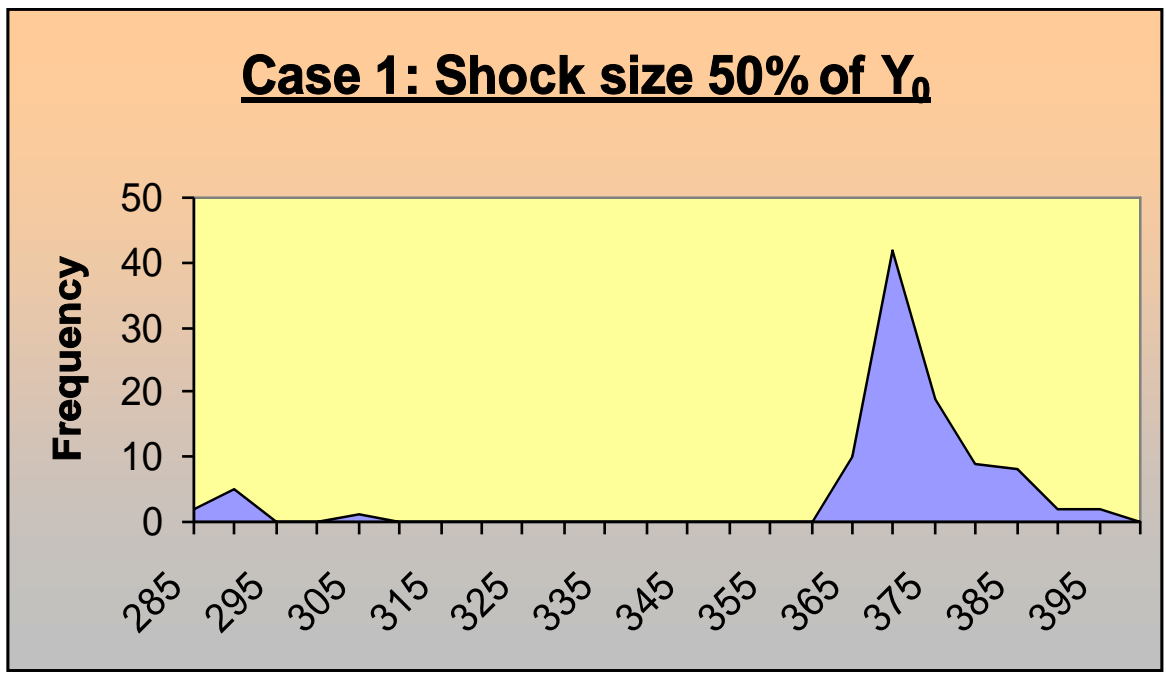

Figure IV: Histogram of simulated fifth generation population sizes

| Table V : Descriptive statistics of fifth generation population sizes | |
|---|---:|
| Mean | 367.72 |
| Standard Error | 1.65 |
| Median | 369.00 |
| Mode | 370.00 |
| Standard Deviation | 16.49 |
| Sample Variance | 272.00 |
| Kurtosis | 24.52 |
| Skewness | -4.75 |
| Range | 117 |
| Minimum | 273 |
| Maximum | 390 |
| Count | 100 |

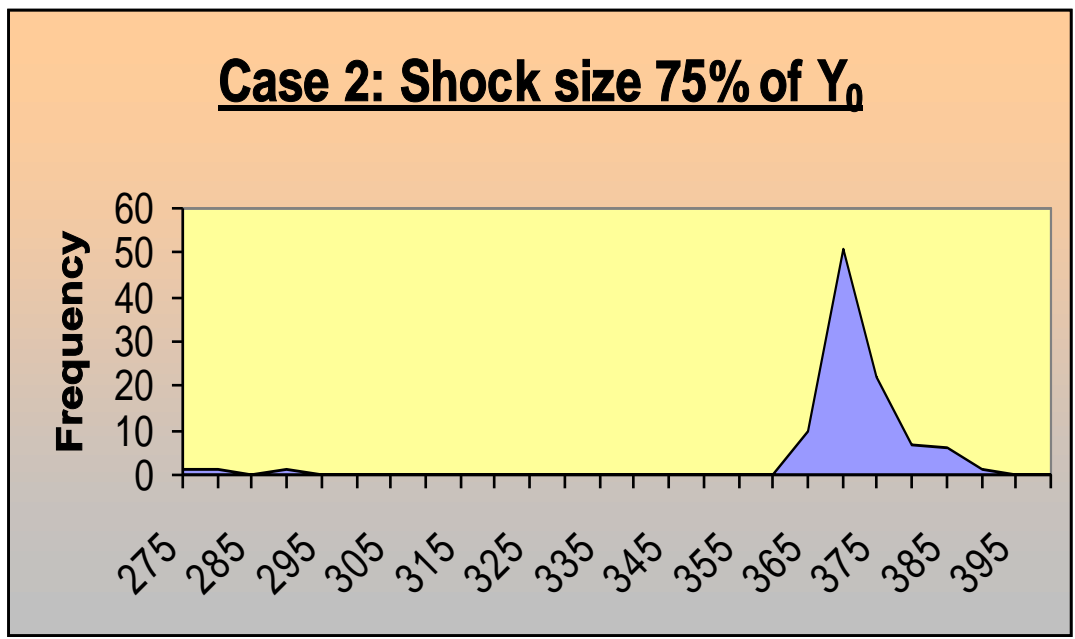

**Case 2: Shock size 75% of $Y_0$**

*Figure V: Histogram of simulated fifth generation population sizes*

| Table VI : Descriptive statistics of fifth generation population sizes | |
|---|---:|
| Mean | 360.25 |
| Standard Error | 4.45 |
| Median | 369.00 |
| Mode | 369.00 |
| Standard Deviation | 44.49 |
| Sample Variance | 1979.52 |
| Kurtosis | 44.44 |
| Skewness | -6.06 |
| Range | 389 |
| Minimum | 0 |
| Maximum | 389 |
| Count | 100 |

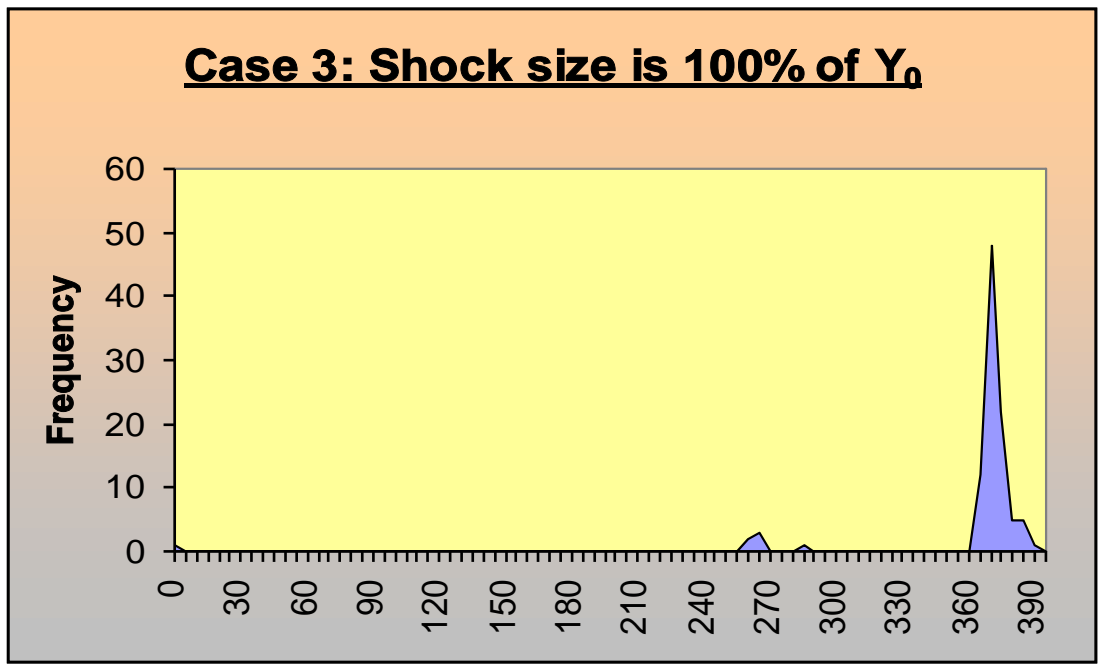

*Figure VI: Histogram of simulated fifth generation population sizes*